\documentclass[manuscript]{acmart}
\AtBeginDocument{%
  }

\usepackage{tabularx}
\usepackage{xcolor}
\usepackage{amsmath}
\usepackage{fvextra}
\usepackage{fancyvrb}
\usepackage{listings}
\usepackage{booktabs}
\usepackage{array}
\usepackage{xcolor}
\usepackage{colortbl}
\usepackage{listings}
\usepackage{xcolor}
\definecolor{codegreen}{rgb}{0,0.6,0}
\definecolor{codegray}{rgb}{0.5,0.5,0.5}
\definecolor{codepurple}{rgb}{0.58,0,0.82}
\definecolor{backcolour}{rgb}{0.95,0.95,0.92}
\lstdefinestyle{mystyle}{
    backgroundcolor=\color{backcolour},   
    commentstyle=\color{codegreen},
    keywordstyle=\color{magenta},
    numberstyle=\tiny\color{codegray},
    stringstyle=\color{codepurple},
    basicstyle=\ttfamily\footnotesize,
    breakatwhitespace=false,         
    breaklines=true,                 
    captionpos=b,                    
    keepspaces=true,                 
    numbers=left,                    
    numbersep=5pt,                  
    showspaces=false,                
    showstringspaces=false,
    showtabs=false,                  
    tabsize=2,
}

\definecolor{lightergray}{gray}{0.9}
\newcommand{\code}[1]{\colorbox{lightergray}{\texttt{#1}}}

\DefineVerbatimEnvironment{promptblock}{Verbatim}{
  breaklines=true,
  breakanywhere=true,
  fontsize=\small
}

\lstdefinestyle{promptstyle}{
    basicstyle=\ttfamily\scriptsize,
    breaklines=true,
    breakatwhitespace=false,
    keepspaces=true,
    columns=fullflexible,
    numbers=none,
    frame=none,
    backgroundcolor=,
    showspaces=false,
    showstringspaces=false,
    showtabs=false,
    xleftmargin=0pt,
    aboveskip=0pt,
    belowskip=0pt
}

\setcopyright{acmlicensed}
\copyrightyear{2026}
\acmYear{2026}
\acmDOI{XXXXXXX.XXXXXXX}

\begin{document}

\title{Remind Me To Check The Stove Before I Leave The House: Authoring Personalized Context-Aware Smart Home Reminders Using Everyday Language}

\renewcommand{\shorttitle}{Authoring Personalized Context-Aware Smart Home Reminders Using Everyday Language}

\author{Reina Szeyi Chan}
\email{chan.szey@northeastern.edu}
\affiliation{%
  \institution{Northeastern University}
  \country{USA}
}

\author{Sujendra Jayant Gharat}
\email{gharat.su@northeastern.edu}
\affiliation{%
  \institution{Northeastern University}
  \country{USA}
}

\author{Maya Lampi}
\email{lampi.m@northeastern.edu}
\affiliation{%
  \institution{Northeastern University}
  \country{USA}
}

\author{Yueran Jia}
\email{jia.yuer@northeastern.edu}
\affiliation{%
  \institution{Northeastern University}
  \country{USA}
}

\author{Avi K Srinivasan}
\email{srinivasan.av@northeastern.edu}
\affiliation{%
  \institution{Northeastern University}
  \country{USA}
}

\author{Xiang Zhi Tan}
\email{zhi.tan@northeastern.edu}
\affiliation{%
  \institution{Northeastern University}
  \country{USA}
}

\renewcommand{\shortauthors}{Chan et al.}

\newcommand*{\reina}{\textcolor{purple}}
\definecolor{rowgray}{RGB}{245, 247, 250}
\begin{abstract}

Reminder systems commonly rely on fixed schedules, location triggers, or simple rules, limiting their ability to leverage the rich sensing capabilities of modern smart homes. A key challenge lies in enabling users to specify context-aware reminders without requiring complex configurations. We present a system pipeline that supports reminder authoring through natural language and conversational interaction. The pipeline translates user requests into structured representations and executable logic, incorporating time-based, activity-based, sensor-based, and state-based conditions. We conducted two studies to examine how users express reminder intent and how conversational support influences the authoring process. In Study 1 (N=40), we analyzed 233 user-authored reminders and identified challenges in expressing reminders with diverse and complex logic. Based on these findings, we refined the system and evaluated it in Study 2 (N=10), demonstrating improved handling of time-based, activity-based, sensor-based, and state-based conditions. Our results highlight the diversity and ambiguity of user expressions and show that conversational guidance can help structure these expressions into flexible, context-aware reminders.
\end{abstract}



 \maketitle

\section{Introduction}

People often need reminders not simply at a fixed time, but at the right moment in the flow of everyday life. A person may want to be reminded to take supplements after dinner, wash their hands when they arrive home from work, or take food out of the microwave if they forget it. These reminders are easy to express in everyday language, yet difficult to create concisely with existing reminder systems. Today, most reminder tools still rely on fixed schedules, singular triggers, or location-based cues~\cite{ludford2006because, graus2016analyzing, draxler2022agenda} and are one-off commands -- ``remind me to call my son at 7pm'' or ``show my grocery list when I am at the grocery store''.  
However, many everyday activities do not follow a fixed schedule. Because time-based reminders must be tied to a specific clock time, they require assumptions about when the relevant activity will occur~\cite{jindal2019heuristic}. As a result, they may fail to capture the complexity and variability of real-world routines.

Smart home technologies provide an opportunity to create reminders that are more closely aligned with everyday activity, rather than relying on users or systems to estimate the right time. For example, voice assistants like Alexa can notify users when someone is at the door, and phones can remind us to drink water when we wake up in the morning.
These devices increasingly assist with routine tasks by sensing, interpreting, and responding to changes in our environment~\cite{wilson2017benefits}. 
Beyond simple event detection, sensing capabilities can enable human activity recognition (HAR)~\cite{thukral2025layout}, offering new opportunities to condition on higher-level activities. With little or no training data, HAR can now detect complex resident activities~\cite{al2020zero, vettoruzzo2025efficient}. Combined with ambient sensor data, these capabilities enable a broader range of context-aware reminder scenarios that go beyond singular events. While these systems capture a wealth of contextual information about daily activities, their potential to support more meaningful, personalized experiences, such as intelligent reminder systems, remains underexplored. The challenge is that useful everyday reminders often depend on context that is richer than a single clock time, location, or sensor event.

The most common approach is trigger-action programming (TAP) (e.g., ``IF [trigger], THEN [action])~\cite{ur2014practical}, where a single event (``front door open'') or a specific time can trigger the reminder. While effective for simple tasks, this model breaks down for reminders that depend on more complex or delayed user behaviors~\cite{chi2022delay, liu2023understanding}. For instance, the reminder -- ``\textit{remind me to take the food out of the microwave if I forget.}'' -- cannot be directly inferred from a single sensor reading. Instead, it requires the system to detect and interpret a sequence of sensors over a period of time. First, it needs to detect that the microwave door has been opened, indicating food was placed inside. Next, it detects a rise and fall in the microwave's power consumption, indicating the microwave was used. Finally, the microwave door needs to remain closed for a fixed duration. Only after that can the system infer that the person has forgotten to take out the food. 

One straightforward approach is to extend existing TAP and rule-based methods by adding additional fields for recognizable activities, options for delays, a process to detect a sequence of sensor readings, and implementation of multi-condition triggers. However, this approach can quickly lead to large and complex configuration files that are difficult for average users to create and manage~\cite{mccall2023towards, garg2019understanding, mennicken2014today}. In addition, sensor availability and deployment vary widely across households, making it challenging to design one-size-fits-all rule sets that generalize well in real-world settings~\cite{thukral2025layout, bouchabou2023smart}. As a result, system capabilities may not be transparent to users, and user intent may be constrained or misinterpreted by the system.

In this work, we propose a conversational reminder authoring pipeline that guides users in creating complex, context-aware reminders that can be triggered by various sensor readings. The core of the system is an alternative representation of reminder execution logic, where each trigger is defined as a function that takes in the time, sensor, and HAR information as input.
However, requiring users to directly construct such function remains challenging, as end-user programming approaches (e.g., IFTTT-style systems) still demand users to translate their intent into formal structures~\cite{ur2014practical,corno2020heytap}.
Rather than requiring users to construct these functions directly, we introduce a conversational pipeline that allows users to describe reminders using everyday language. 
Prior work has shown that natural language provides an intuitive way for users to express reminder intent in terms of routines and situations~\cite{king2024sasha, clark2016towards,corno2020heytap}.
Building on this, our pipeline translates these conversations into structured function representations. The system uses a large language model (LLM)-powered chat assistant to guide users through the authoring process, helping them refine underspecified requests and align their intent with the capabilities of the smart home environment.

While this approach aims to make reminder authoring more natural and accessible, designing such a system is challenging. Users may express the same reminder intent in many different ways, and the system must handle this variability while supporting flexible interaction. Moving beyond simple, single-trigger reminders introduces additional challenges, and how users author more complex reminders in a conversational setting remains underexplored.
To address this, we also investigate how users express reminder intent in natural language, how these expressions align with system capabilities, and how conversational systems can support the creation of context-aware reminders. To this end, we conducted two studies across six everyday scenarios. In Study 1, we collected 233 reminders from 40 participants and analyzed reminder authoring interactions to understand how users describe reminders and to identify gaps between user intent and system capabilities.
These findings allowed us to refine the authoring pipeline and better align it with how users naturally create reminders. The updated system was then evaluated with 10 new participants.

Our results show that users express reminder intent in diverse and often underspecified ways, and that system-guided conversational support can help structure these expressions into feasible, context-aware reminders. By better understanding how users author reminders through conversation, we were able to improve the pipeline's accuracy from 45.5\% to 76.7\%, showing it can handle time-based, activity-based, sensor-based, and state-machine based reminders. Our work makes the following four contributions:
\begin{enumerate}
    \item A function representation for context-aware reminder triggers that supports combinations of time-based, activity-based, sensor-based, and state-machine based reminders.
    \item A conversational reminder authoring pipeline that maps natural language input to structured representations and executable logic.
    \item An empirical study of how users express reminder intent and how these expressions align with the capabilities of a smart home system.
    \item Design insights for supporting flexible, context-aware reminder authoring through conversational interaction.
\end{enumerate}

\section{Related Work}
\subsection{Reminders as Memory Aids}
Reminders serve as cognitive offloading tools that help people manage memory demands in daily life~\cite{gilbert2020optimal, risko2016cognitive}. They reduce the burden of remembering future intentions by externalizing information, such as buying groceries~\cite{pradhan2020use}, daily home activities~\cite{lariviere2021placing, thomas2014verbal}, taking medications~\cite{stawarz2014don, stawarz2016understanding}, or even logging social interactions~\cite{ramos2016designing}. Reminders have become deeply embedded in people’s everyday routines.

Creating a reminder typically involves specifying when and under what conditions an intended action should occur~\cite{chaminda2012smart,dey2000cybreminder}. Prior work identifies several common types of reminder triggers: time-based~\cite{brewer2017remember, hicks2005task}, event-based~\cite{smith2010cognitive, hicks2005task}, and location-based~\cite{ludford2006because, sohn2005place}. Each type requires users to define key components such as the intended action (often phrased as a predicate statement), the trigger condition (time or event), and in some cases, additional contextual parameters~\cite{graus2016analyzing}. These components form the foundation of how reminders are authored and activated.

Traditionally, these reminders have been implemented using physical memory aids like Post-its, calendars, or handwritten notes~\cite{dey2000cybreminder}. These tools are often placed in visible or contextually relevant locations, such as a fridge door or a work desk, to maximize the chance of being noticed at the right time and in the right place. However, even well-placed reminders can fail: they may be misplaced, overlooked, or not seen when most needed~\cite{raghunath2020creating}. For instance, people might forget to bring a written grocery list to the store, undermining the reminder’s intended purpose. With the advancement of personal and ubiquitous technologies, reminders have moved from physical forms to digital platforms. Today, reminders are commonly delivered through mobile calendars, to-do list applications, and smartwatch notifications across various contexts and devices~\cite{gurol2013mobile, stawarz2014don,ludford2006because}. In addition to supporting basic time- and event-based prompts, reminders have increasingly been used to encourage behavior change and habit formation~\cite{stawarz2015beyond}. For example, just-in-time reminders to intervene in smartphone overuse, encouraging healthier digital habits~\cite{hiniker2016mytime, ko2016lock, okeke2018good}. For populations with more complex daily routines, such as older adults managing medication schedules and household tasks, reminders play an important role in supporting independence and adherence to everyday activities~\cite{mathur2022collaborative, chan2025insights}.

Despite these advances, many digital reminder systems still rely on single-cue triggers that require manual and precise configuration. They often fail to adapt to dynamic routines or nuanced contextual factors that affect when and how people act on their intentions. This motivates our work, which explores more flexible, context-aware reminders that adapt to users’ changing environments and behaviors.

\subsection{Smart Home Technologies for Contextual Reminder Support}
Smart home technologies and Internet of Things (IoT) devices offer new ways to trigger reminders by providing continuous streams of contextual data. Sensors such as motion detectors, door contacts, smart plugs, and connected appliances can capture fine-grained information about users’ activities and environments~\cite{satyanarayanan2002pervasive}. Prior work has explored how these data streams can enable personalized notifications, just-in-time prompts, and adaptive interruption management~\cite{ho2005using, nahum2016just}.

Within this ecosystem, voice-controlled smart assistants (e.g., Amazon Alexa, Google Assistant, Apple Siri) have made it easier to create reminders through hands-free, natural language commands~\cite{shade2020voice, pradhan2020use}. These tools reduce interaction effort and increase accessibility. At the same time, smart home products support embedded reminder-like functionality. For example, smart plugs may automatically turn off appliances after inactivity, and smart lights may activate according to established routines~\cite{zeng2017end}. These features integrate reminders into the physical environment, aligning with users’ habits and enabling more seamless interactions.

However, existing reminder systems, including those powered by smart home devices, still rely on static or rule-based triggers, such as specific times or locations, configured in advance~\cite{nandi2016automatic, anik2025programming}. These predefined rules often do not reflect the evolving and dynamic nature of daily life, leading to reminders that may fire too early, too late, or not at all. To address these limitations, our work builds on the sensing capabilities of smart homes to support more adaptive and context-aware reminders. By providing users with a basic understanding of available sensor capabilities during reminder creation, our system enables them to set personalized reminders based on contextual conditions, rather than relying solely on fixed schedules.

\subsection{End-User Programming and LLM for Automations}
Prior work offers two main approaches to creating home automations~\cite{song2024learning}: activity prediction and TAP. Activity Prediction learn from historical sensor streams to predict the likelihood of performing the activity in the future~\cite{minor2017learning}. TAP allows end users to encode rules to automate systems~\cite{ur2016trigger, zhang2019autotap}. 

TAP has gained widespread adoption due to its ability to provide users with direct control over their automations~\cite{ghiani2017personalization}. It empowers individuals without programming experience to configure smart environments and tailor them to their personal routines. A common approach for expressing such rules is the ``IF [trigger], THEN [action]'' paradigm~\cite{ur2016trigger}, which has been popularized by platforms such as IFTTT~\cite{ifttt}, Zapier~\cite{zapier}, Home Assistant~\cite{homeassistant}, and Microsoft Power Automate~\cite{powerautomate}. However, TAP fundamentally relies on singular trigger-action pairs, which limits its ability to represent more complex, context-dependent behaviors~\cite{brackenbury2019users}. In practice, however, many everyday behaviors may involve more than one condition~\cite{mi2017empirical, huang2015supporting}. While some systems allow users to chain multiple rules or conditions, this approach adds complexity for users~\cite {huang2015supporting, zhang2020trace2tap}. As the number of rules increases, configurations become harder to manage and reason about, and can lead to unintended behaviors such as conflicting rules, wrong triggers, or infinite loops~\cite{mi2017empirical, brackenbury2019users, nacci2018buildingrules,brich2017exploring}. These limitations highlight the need for alternative approaches that can move beyond singular trigger-action representations and better support complex, real-world scenarios.

One approach is the use of natural language as an interface for automation. Prior work has demonstrated that natural language can be translated into executable automation~\cite{liu2023understanding,corno2020heytap}. For instance,~\citet{yusuf2022accurate} proposed RecipeGen, where it identifies trigger and action from natural language descriptions by using a Transformer sequence-to-sequence architecture. Large language models (LLMs) extend this capability by enabling direct conversion of user intent into machine-interpretable commands, offering an alternative to traditional automation methods. ~\citet{gallo2023towards} proposes a conversational agent that combines ChatGPT and Rasa to help create trigger-action rules for users. Additionally, ~\citet{king2024sasha} explores using LLMs to respond to under-specified user commands by generating control and automation routines for smart home devices. Building on this direction, our system introduces a chatbot-based reminder interface that leverages LLMs. The chatbot converts natural language input to executable reminder logic while considering the available home sensor infrastructure and its feasibility. This approach removes the need for users to manually encode rules and supports flexible, real-time, and context-aware reminder creation through natural language.

\section{Methodology}

Our work consisted of two parts. As there has been limited prior work on understanding how users author reminders beyond singular, time-based trigger in a conversational setting, we first conducted a formative study (Study 1) to collect rich data on how users naturally express reminder intent. This formative study utilizes our initial reminder authoring system as participants would need to interact with a working conversational systems. The goal of Study 1 was not to evaluate system performance but understand how users describe reminders in natural language through a conversational interface and to identify gaps between user expectations and the system’s current capabilities in the reminder-authoring pipeline. The data collected in this study was used to refine the reminder authoring pipeline and improve the LLM prompting strategy. Based on the insights gained from Study 1, we then conducted a smaller follow-up study (Study 2) with new participants to evaluate the refined system. The two studies followed similar procedures, with the primary differences being the design of the reminder authoring system and an updated set of human activity recognition and sensor labels.

\section{Study 1: Formative Exploration of User Reminder Authoring}
For Study 1, we used our initial version of the reminder authoring pipeline to collect participant data to understand their conversational patterns and refine our pipeline.
In this section, we begin by describing that pipeline. We then outline the study design and procedure, followed by the reminder scenarios used in the study. Next, we analyze the collected data, including the types and categories of reminders authored by participants. We also examine common breakdowns between user intent and system output, alongside aspects of the system that worked well. Finally, we discuss how these findings informed subsequent improvements to the reminder authoring pipeline.

\subsection{System Overview}
\begin{figure*}
  \includegraphics[width=\textwidth]{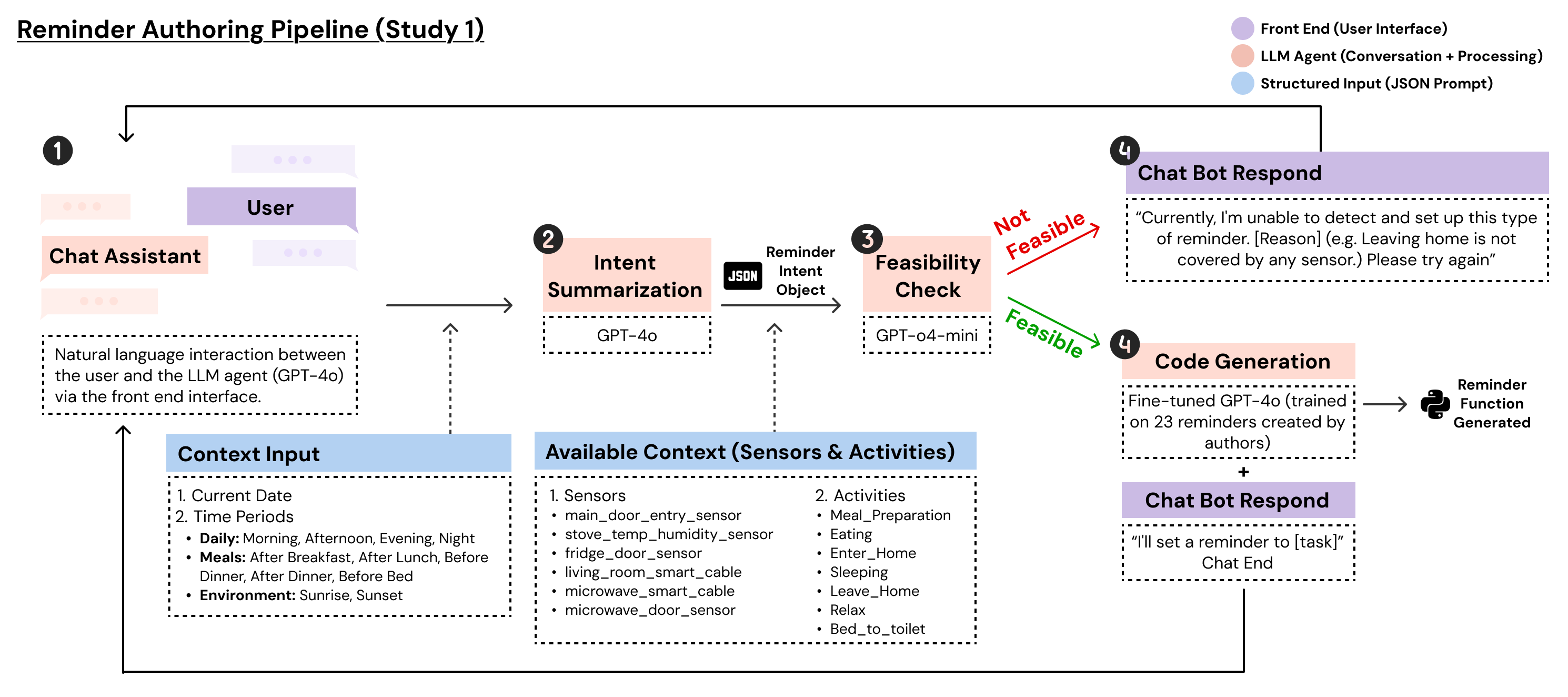}
  \caption{An overview of the reminders authoring pipeline for the reminder triggers used in Study 1. This is where user input, intent extraction, and reminder generation occur.}
  \label{fig:system-study-1}
  \Description{An overview of the reminders authoring pipeline for the reminder triggers used in Study 1. This is where user input, intent extraction, and reminder generation occur.}
\end{figure*}

In the section, we describe the reminder authoring pipeline used in study 1, an overview of the system is shown in Figure~\ref{fig:system-study-1}. The users can create reminders using natural language through a chat interface. These reminders are then translated into structured, executable functions by the authoring pipeline. We first describe the representation of the generated reminders, followed by the pipeline that produces them.

\label{sec:authoring_reminders}
\subsubsection{Reminder Trigger Representation}
To support complex conditions, our system represents each reminder trigger as a Boolean Python function that determines whether the reminder should be triggered. Each function takes the current time, sensor state data, human activity recognition data, and a shared memory structure (blackboard) as input. 
The trigger functions are evaluated at fixed intervals, and the blackboard allows information to be shared across evaluation.
We support human activity labels based on the CASA datasets~\cite{cook2012casas} , which provides information about detected user activities at the time of invocation. 
The sensor data records the current state of household smart home devices, including contact sensors (True if open), motion sensors (True if there is motion), and power sensors (numeric amperage value).
The trigger function returns Boolean value (True or False) indicating whether the reminder conditions have been met, which is then used to trigger the reminder. The expressiveness of this function-based representation enables the construction of triggers with multiple conditional statements and complex logic. The code below shows an example that defines a reminder that notifies the user when the front door is opened while the stove is still on, by combining sensor state with historical context stored in the blackboard to detect state changes.

\begin{lstlisting}[language=Python]
def reminder(time=None, activity_data=None, sensor_data=None, blackboard=None):
  s = sensor_data or {}
  bb = blackboard if isinstance(blackboard, dict) else {}
  door_open = bool(s.get("contact_front_door"))
  prev_door_open = bb.get("prev_door_open", False)
  opening_edge = (not prev_door_open) and door_open
  stove_on = bool(s.get("plug_kitchen_counter_1")) or bool(s.get("plug_kitchen_counter_2"))
  bb["prev_door_open"] = door_open
  return opening_edge and stove_on
\end{lstlisting}

The trigger function is part of a reminder intent object (stored as a JSON file), which also includes the reminder message, whether the reminder is recurring, and whether it should occur only on a specific date. 
The manual generation of this configuration by users would be challenging. However, by leveraging the code generation and conversational capabilities of LLMs, users can generate these complex configurations more easily. In the following subsections, we describe the pipeline to generate the trigger function and reminder configuration file.

\subsubsection{LLM-Based Conversational Interface for Reminder Creation}
Users interact with a chat interface to create reminders using natural language. The chat assistant is powered by an LLM and guided by a structured prompt that enforces concise, context-aware dialogue for efficient reminder creation. In this step, we leverage the GPT-4o model as the underlying LLM. The prompt frames reminder authoring as a structured conversational process in which the assistant incrementally collects the required information needed to define a reminder. This process focuses on two required components: \code{WHAT}, which specifies the task to be remembered, and \code{WHEN}, which specifies when the reminder should occur.
\code{WHEN} is not limited to time and can be any reasonable event or activity at home (e.g., ``when I get home'' or ``after dinner'').
Additional optional slots include \code{DATE}, which can be specified as ``today,'' ``tomorrow,'' or a specific calendar date, and \code{RECURRENCE}, which defines whether the reminder occurs once or follows a repeated schedule (e.g., daily or weekly). Since reminders cannot be scheduled without both \code{WHAT} and \code{WHEN}, the chat assistant ensures that these required slots are filled before proceeding. 
The chat assistant follows a constrained interaction policy designed to minimize user effort. It asks only for missing information and limits itself to one follow-up question at a time. 
When the user has not specified when the reminder should occur (e.g., ``remind me to water the plant.''), the assistant asks a concise question to obtain this information. 
Once all required details are collected, it restates the complete reminder and asks the user for confirmation. After receiving explicit confirmation, the assistant triggers the next stage of the reminder pipeline.

\subsubsection{Intent Summarization}
In the next stage of the pipeline, an intent module converts the user's request into the reminder intent object. The module is also given user-specific context inputs.
These include the current date, predefined time periods (e.g., morning, afternoon, evening, night), user time preferences (e.g., before or after meals, before bed), and environmental cues (e.g., sunrise, sunset).
These contextual inputs are used when the user's request is sometimes underspecified or expressed in relative or abstract terms. For example, expressions such as ``in the evening,'' ``before dinner,'' or ``around bedtime'' require the system to map user language to concrete time-based or activity-based representations. By incorporating these context signals, the system can infer appropriate trigger conditions and generate a more complete and executable reminder intent.

We leverage the GPT-4o model as the underlying LLM here as well. The module extracts exactly one reminder that includes both the \code{WHAT} and \code{WHEN} components, and represents it as a complete sentence beginning with ``Remind me to ...'' In addition to the task, the module filled in structured fields for \code{DATE}, \code{TIME} (which includes both exact clock times, when available, and inferred time expressions derived from natural language phrases such as ``after dinner''), \code{RECURRENCE} (which specifies whether the reminder occurs once or follows a repeated schedule, such as daily or weekly), and \code{PRIORITY} (which indicates the importance level of the reminder, categorized as high, medium, or low). 
There are also a set of deterministic rules defined in the prompt. For example, If no details are provided, the module uses sensible defaults: ``today'' for the date, a one-time reminder for recurrence, and ``medium'' for priority. Times are converted to 24-hour format when possible, while vague expressions like ``after dinner'' are kept as inferred times.

\subsubsection{Feasibility Check}
Our pipeline then checks if a reminder can be reliably triggered using available sensing capabilities. We leverage the GPT-4o-mini model as the underlying LLM for this module. Given a structured reminder sentence, the module evaluates the detectability of the \code{WHEN} component based on predefined rules and the set of available sensors and activities in the environment. Time-based triggers are always considered possible. The module checks whether the expression corresponds to a detectable activity (e.g., entering the home) or a detectable sensor event (e.g., opening a device), and verifies that the associated location is instrumented with at least one sensor. 

The module also applies constraint rules to handle unsupported cases. For example, triggers phrased as ``before <activity>'' are considered non-detectable, as the system cannot reliably predict the start of an activity without an explicit signal, reflecting limitations that also exist in the current sensing capabilities.
If no valid trigger can be identified, the reminder is marked as \textit{not feasible}. The module then outputs a brief explanation and returns to the chat phase of the interaction.

\begin{figure*}
  \includegraphics[width=0.9\textwidth]{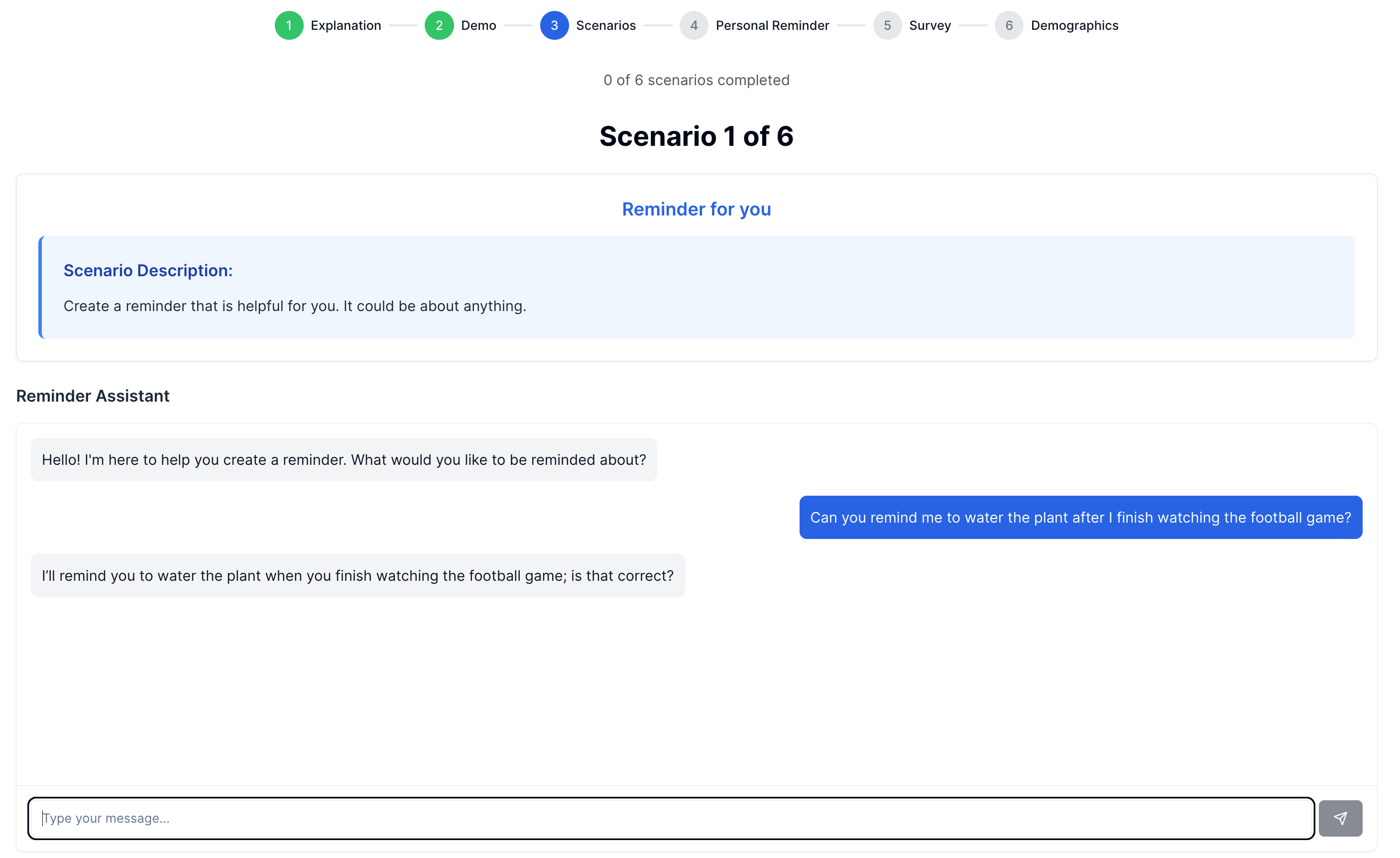}
  \caption{Example of the chat-based reminder authoring interface used by participants during the study.}
  \label{fig:screenshot}
  \Description{Example of the chat-based reminder authoring interface used by participants during the study.}
\end{figure*}

\subsubsection{Code Generation}
Once a trigger is deemed feasible, the pipeline generates the corresponding trigger function. The prompt includes predefined detectable activities and sensor information to guide the code generation of condition checks. As this task differs significantly from standard programming tasks at that time, we used a fine-tuned (with 23 representative reminders) GPT-4o model for the generation task. The code generation process operates in parallel with the conversational interaction. While the reminder function runs in the background using the fine-tuned model, the chat assistant immediately provides a confirmation response and ends the conversation, keeping the interaction responsive without waiting for code generation to complete.

\begin{table}[t!]
\centering
\small 
\renewcommand{\arraystretch}{1.6}
\begin{tabular}{p{0.6cm} p{13.8cm}}
\toprule
\multicolumn{2}{l}{\bfseries Descriptions of Scenarios for Reminder Creation} \\
\midrule
1 & Create a reminder that is helpful for you. It could be about anything. \\
\rowcolor{rowgray}
2 & Imagine you're John, a busy professional who enjoys watching your favorite TV show during dinner. You usually heat your meal in the microwave, but often become absorbed in other tasks or distracted by your show and forget your food until it's cold.\\
3 & Imagine you're Emma, a retired schoolteacher who enjoys cooking breakfast in your cozy kitchen. Your mornings are usually busy with errands, visits, or volunteer activities. Recently, you've noticed feeling uneasy after leaving home, worrying whether you've left the stove on.\\
\rowcolor{rowgray}
4 & Imagine you're Emily, a thoughtful daughter who often visits your mother, Sarah. Knowing she frequently skips breakfast during busy mornings, you cooked her favorite meal last night and placed it in the fridge for her to enjoy in the morning. You don't want to disturb her sleep with calls or texts, but you're concerned she might forget about the breakfast.\\
5 & Imagine you're Rachel, a healthcare worker who spends your days caring for patients, often returning home tired after long shifts. Sometimes, you're so exhausted or distracted upon arriving home that you unintentionally skip important routines, like washing your hands.\\
\rowcolor{rowgray}
6 & Imagine you're Margaret, a retired senior who enjoys peaceful evenings at home reading, watching TV, or chatting with family. Each evening after dinner, you need to take supplements important for your health, but lately you've sometimes forgotten, causing you worry at bedtime.\\
\bottomrule
\end{tabular}
\caption{Scenario descriptions used in the reminder creation task.}
\label{tab:scenario}
\end{table}

\subsection{Study Design}
Our formative study was a web-based experiment that took approximately 20 minutes to complete. Participants were recruited through Prolific and received \$6 USD for participation. Recruitment was limited to individuals located in the United States, and the sample was balanced by gender. The study protocol was approved by the university’s Institutional Review Board (IRB).

After providing informed consent, participants completed a tutorial that introduced the chat-based interface used to create reminders, during which they were guided through the steps to submit a reminder request. Participants then proceeded to the main task, where they created reminders across six scenarios using the chat-based interface (Figure~\ref{fig:screenshot}). The first scenario was open-ended and always presented first, allowing participants to create a reminder of their choice. The remaining five scenarios were presented in randomized order to reduce ordering effects. After completing all tasks, participants filled out the System Usability Scale (SUS) questionnaire~\cite{brooke1996sus} using a five-point Likert scale. Finally, participants provided demographic information before being redirected back to Prolific for payment.

\subsubsection{Scenarios}
We designed six scenarios to represent common situations in which people use reminders in daily life (Table~\ref{tab:scenario}). These scenarios cover a range of use cases, including safety, health, planning, and routine activities. The first scenario was open-ended to capture participants’ natural reminder needs, while the remaining scenarios provided structured contexts to ensure consistency across participants. Each scenario prompted participants to describe a reminder through a conversational interaction with the system.

\subsection{Method}
\subsubsection{Analysis Method}
We analyzed the collected data using qualitative content analysis~\cite{hsieh2005three, mayring2021qualitative} with an iterative categorization scheme. First, the first author reviewed all sessions to become familiar with the dataset and identify recurring patterns in how participants expressed reminder intent. Based on this review, we developed an initial set of categories with definitions and examples. Next, two authors independently applied the initial categories to the full dataset. To refine the categorization scheme, all three authors jointly reviewed a randomly selected subset of 33 sessions, discussed disagreements and clarified category definitions. The category scheme was then updated to reflect these refinements. Finally, the two coders revisited the remaining 200 sessions and recoded them using the updated scheme to ensure consistent application. Any remaining disagreements were resolved through discussion until consensus was reached. The finalized categories were then used to derive quantitative summaries, including success rates and the distribution of reminders across reminder trigger types and functional purposes.

\subsubsection{Categorization Scheme}
To analyze the reminders created by participants, we categorized each session along three dimensions: reminder trigger type, functional purpose of the reminder, and correctness of the generated reminder output. Categorization was based on the final user intent expressed in the conversation session. Sessions marked as incomplete were excluded from analysis.

Reminders were first categorized according to their primary triggering mechanism. We identified four types of triggers: 
\begin{enumerate}
    \item \textbf{Time-based triggers} are scheduled at a specific time or date and are triggered by explicit time related conditions. If a reminder included a specific time as its primary trigger, it was classified as time-based. 
    \item \textbf{Activity-based triggers} are triggered by recognizable human activities such as eating, leaving home, or relaxing. 
    \item \textbf{Sensor-based triggers} are triggered by environmental sensor events or states, such as detecting that a door has been opened or that a device is powered on. 
    \item \textbf{State machine triggers} capture sequences of sensor events and rely on reasoning about event order and history rather than a single sensor state. For example, a reminder may depend on a sequence in which the microwave door is opened, the microwave is powered on, then powered off, and the door remains closed. When multiple conditions were present in a reminder request, reminders were classified according to the primary triggering mechanism, while additional conditions such as time, activity, or sensor constraints were treated as secondary conditions.
\end{enumerate}

In addition to trigger type, reminders were also categorized by functional purpose. We identified four primary reminder categories:
\begin{enumerate}
    \item \textbf{Habit reminders} refer to recurring routine behaviors, such as drinking water regularly or taking vitamins after dinner. 
    \item \textbf{Appointment reminders} refer to commitments scheduled at a specific time that often involve coordination with others, such as meetings or doctor visits. 
    \item \textbf{One-time reminders} include reminders for tasks that need to be completed once or prepared ahead of time, such as taking food out of the refrigerator for breakfast or preparing items for the following day.
    \item \textbf{Alert reminders} are designed to detect potential oversights or safety concerns and typically involve conditional logic, such as reminding the user if a door is left open.
\end{enumerate}

Finally, we evaluated the correctness of the code output generated by the system. Each generated reminder was labeled as correct, partially correct, or incorrect. A reminder was labeled correct if the generated output fully matched the user’s intent. A reminder was labeled partially correct if it captured the intended task but contained specific errors. These errors included referencing sensors not in the supported sensor set, specifying incorrect time conditions, using incorrect activity triggers, or implementing incorrect logic in reminders that require state-machine reasoning. Reminders were labeled incorrectly if the generated output was unrelated to the user’s request. Final accuracy was calculated based only on correctly generated code outputs.

The reminder generation system operated using a predefined set of sensors and activity labels available in the smart home environment based on the CASAS dataset~\cite{cook2012casas}. In our study, the activities used included meal preparation, eating, entering and leaving the home, sleeping, and relaxing. The supported sensors included a main door entry sensor, a stove temperature and humidity sensor, a refrigerator entry sensor, a living room smart cable, a microwave smart cable, and a microwave door entry sensor. Generated reminders that referenced sensors outside this predefined set were considered incorrect.

\subsection{Results}
\subsubsection{Participant Demographic Information}
In this study, 40 participants (20 male, 19 female,  and 1 non-binary) were recruited. Participants ranged in age from 23 to 65 years ($M = 43.15$, $SD = 10.49$). Participants reported moderate to high LLM usage frequency ($M = 5.18$, $SD = 1.32$ on a 7-point scale, where 7 indicates daily usage), with 67.5\% indicating frequent use (ratings of 5--7).

\subsubsection{System Usability Scale (SUS) Results}
Participants completed the System Usability Scale (SUS) after each study. In Study 1, the interface received a mean SUS score of 79.81 (SD = 15.86, $n=40$), suggesting good usability with challenges for some participants~\cite{bangor2008empirical, sauro2016quantifying}.  

\begin{figure*}
  \includegraphics[width=\textwidth]{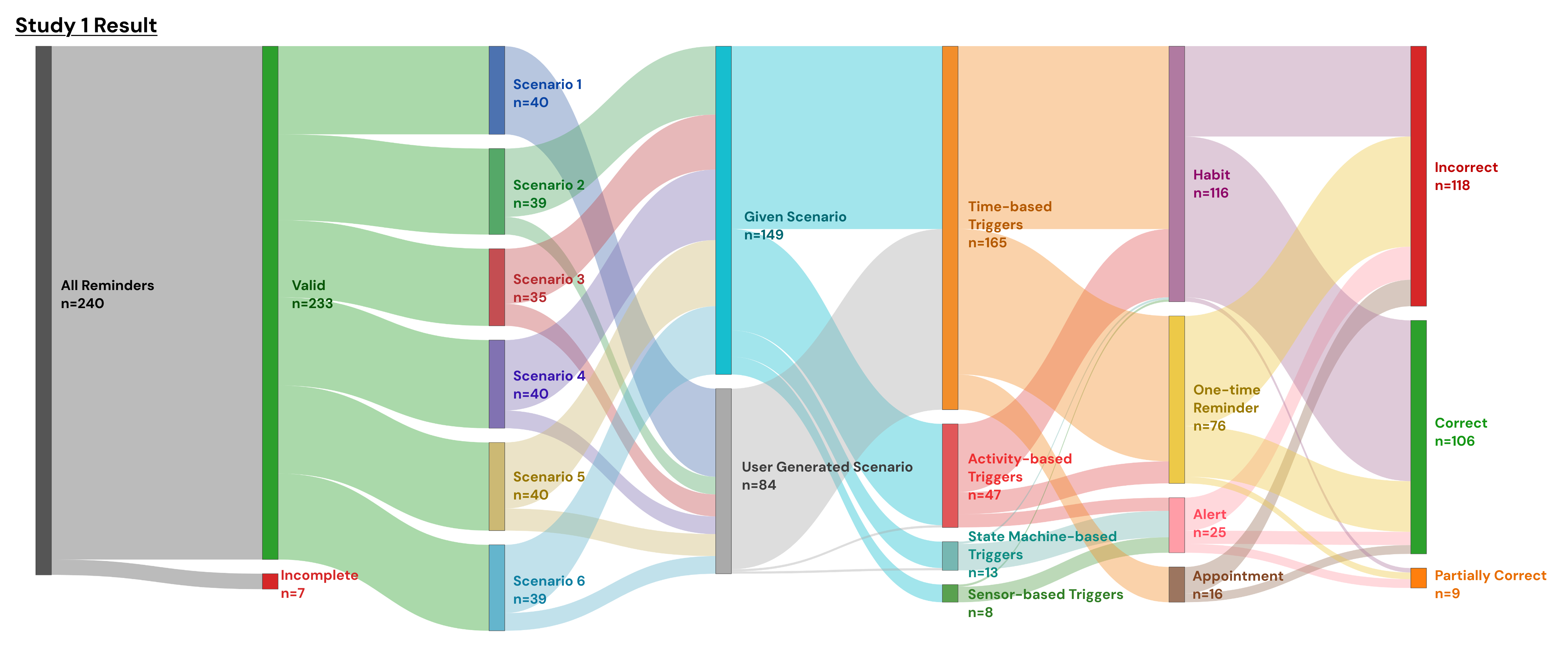}
  \caption{Study 1 results overview. Sankey diagram illustrating the distribution of results across categories. Reminders are first grouped by input validity (valid or incomplete), scenario, and categorized as either given or user-generated scenarios, and then mapped to reminder trigger types and purpose categories. Final outcomes reflect the correctness of the generated code.}
  \label{fig:study1_result}
  \Description{Study 1 results overview. Sankey diagram illustrating the distribution of results across categories. Reminders are first grouped by input validity (valid or incomplete), scenario, and categorized as either given or user-generated scenarios, and then mapped to reminder trigger types and purpose categories. Final outcomes reflect the correctness of the generated code.}
\end{figure*}

\begin{figure*}
  \includegraphics[width=\textwidth]{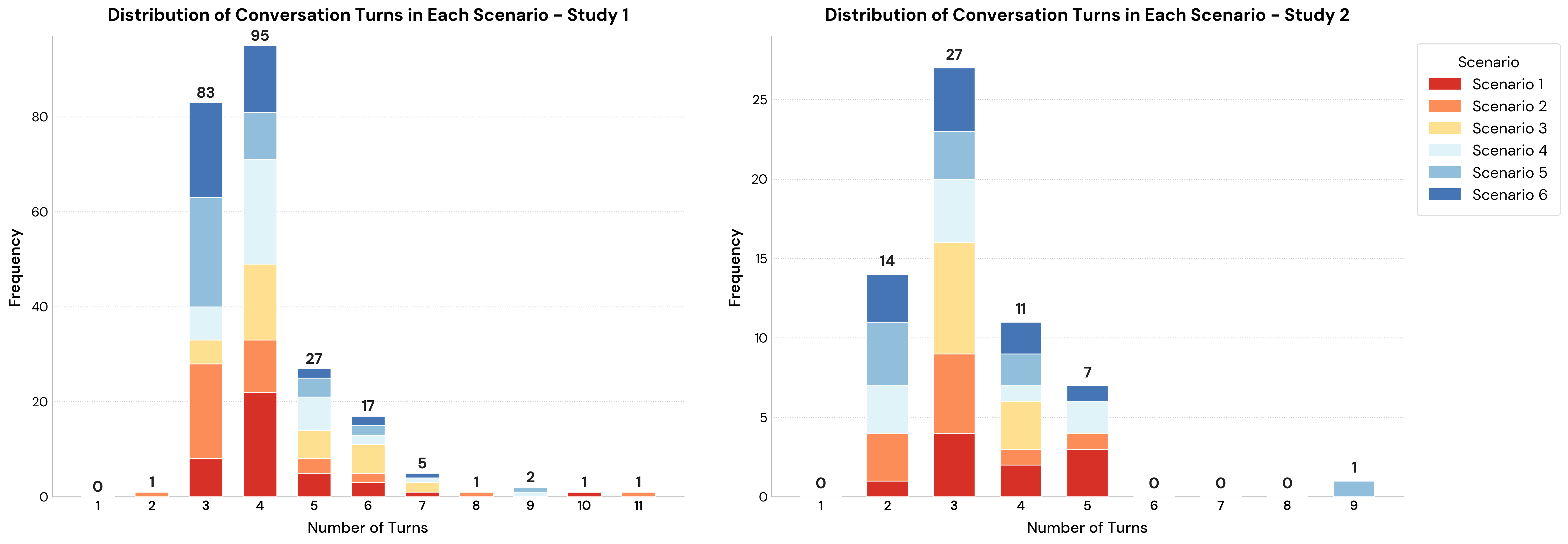}
  \caption{Distribution of the number of conversation turns per interaction across scenarios in Study 1 (left) and Study 2 (right). A conversation turn consists of one user message and one assistant response. Bars are stacked by scenario, and labels indicate the total frequency for each turn count.}
  \label{fig:conversation_turns}
  \Description{Distribution of the number of conversation turns per interaction across scenarios in Study 1 (left) and Study 2 (right). A conversation turn consists of one user message and one assistant response. Bars are stacked by scenario, and labels indicate the total frequency for each turn count.}
  \label{fig:number_of_turns}
\end{figure*}

\subsubsection{Overall Scenarios Result}
In Study 1, 40 participants each completed six reminder authoring scenarios, resulting in 240 reminder creation sessions. The interface limited each scenario to two attempts. In seven sessions, participants reached this limit, and these sessions were excluded from analysis. We analyzed the remaining 233 valid sessions using the iterative categorization scheme.
Participants created reminders based on six predefined scenarios (Scenarios 1-6). However, in some cases (Scenarios 2–6), participants did not follow the given instructions and instead created their own reminders. These were labeled as \textit{user-generated scenarios}. Figure~\ref{fig:study1_result} illustrates the end-to-end distribution of reminders across collection, scenario assignment, reminder type, category, and correct code generation outcome. On average, participants took $4.08$ turns ($SD = 1.27$, $n = 233$) when constructing the reminders (shown in Figure~\ref{fig:number_of_turns}).

Across all sessions, time-based trigger were the most common (n=165). Activity-based trigger accounted for 47 instances, while state machine-based (n=13) and sensor-based trigger (n=8) were less frequently used. Reminders were further categorized into four reminder categories: habits (n=116), one-time reminders (n=76), alerts (n=25), and appointments (n=16). Habit-based reminders constituted the largest category, followed by one-time reminders. In the following, we will dive deeper into \textbf{user-generated scenarios} (n=84) and \textbf{prompted reminder scenarios}, providing a more detailed analysis of category-specific distributions. Table~\ref{tab:study1_study2_scenarios} shows the distribution of reminders and code generation success rate across scenarios.

\subsubsection{User-Generated Scenario Reminder Creation Result}

The first scenario allowed participants to create any reminder they found useful before beginning the structured experimental tasks. This open-ended task helped us understand the types of reminders users naturally want to set in their daily lives. In addition, some participants did not strictly follow the instructions in later scenarios and instead created their own reminders. Because the goal of this study was not to evaluate system performance but to understand how users express reminder needs, we included these reminders in our analysis.

\begin{table}[t!]
\centering
\footnotesize
\renewcommand{\arraystretch}{1.4}
\begin{tabular}{lcccc}
\toprule
 & \multicolumn{2}{c}{\textbf{Study 1 (n=233; 45.5\%)}} & \multicolumn{2}{c}{\textbf{Study 2 (n=60; 76.7\%)}} \\
\cmidrule(lr){2-3} \cmidrule(lr){4-5}
\textbf{} & \textbf{Given Scenario} & \textbf{User Generated Scenario} & \textbf{Given Scenario} & \textbf{User Generated Scenario} \\
\midrule
Scenario 1 & N/A  & 40 (55.0\%)& N/A & 10 (90.0\%)\\
\rowcolor{rowgray}
Scenario 2 & 31 (12.9\%) & 8 (12.5\%)  & 9 (77.8\%) & 1 (0.0\%) \\
Scenario 3 & 25 (48.0\%) & 10 (40.0\%) & 8 (87.5\%) & 2 (50.0\%) \\
\rowcolor{rowgray}
Scenario 4 & 32 (21.9\%) & 8 (50.0\%)  & 8 (100.0\%) & 2 (50.0\%) \\
Scenario 5 & 30 (76.7\%) & 10 (40.0\%) & 8 (75.0\%)& 2 (100.0\%) \\
\rowcolor{rowgray}
Scenario 6 & 31 (71.0\%) & 8 (37.5\%)  & 9 (44.4\%)& 1 (100.0\%)\\
\midrule
\textbf{Total} & 149 (45.6\%) & 84 (45.2\%) & 42 (76.2\%) & 18 (77.8\%)\\
\bottomrule
\end{tabular}
\caption{Distribution of reminders and code generation success rates across scenarios in Study 1 and Study 2. Values represent the number of reminders in each scenario, with percentages indicating the proportion of reminders that resulted in correct code generation.}
\label{tab:study1_study2_scenarios}
\end{table}

\begin{table}[t!]
\centering
{\footnotesize
\renewcommand{\arraystretch}{1.5}
\begin{tabular}{p{3.5cm} >{\centering\arraybackslash}p{2cm} >{\centering\arraybackslash}p{2cm}  p{6.5cm}}  
\toprule
\textbf{Reminder Type} & \textbf{Study 1 (n=84)} & \textbf{Study 2 (n=18) } & \textbf{Example of Reminders} \\
\midrule
House Chores & 22 & 2 & Take out the trash every Tuesday night at 7 PM \\
\rowcolor{rowgray}
Personal Obligations & 17 & 4 & Take my car to get it fixed \\
Medical & 10 & 5 & Go to doctor's appointment tomorrow \\
\rowcolor{rowgray}
Food Preparation & 10 & 2 & Start thawing out dinner at 5 PM today \\
Personal Care and Wellness & 11 & 0 & Log my steps into my walking apps at 6pm every night \\
\rowcolor{rowgray}
Activity with Others & 8 & 1 & Meet up with a friend at 8:30 tomorrow \\
Pet-related Reminders & 6 & 4 & Take cat to the vet on Wednesday \\
\bottomrule
\end{tabular}
}
\caption{Distribution of user-generated reminder types across Study 1 and Study 2. Values indicate the number of reminders in each category, with representative examples shown.}
\label{tab:reminder_types}
\end{table}

In total, we collected 84 user-generated reminders, including 40 from Scenario 1 and 44 from other scenarios in which participants chose to create their own reminders rather than follow the given instructions. 
Table~\ref{tab:reminder_types} shows the distribution of reminder type of the user-generated reminder.
Through qualitative analysis, we identified seven common types of reminders: house chores (n=22), personal obligations (n=17), personal care and wellness (n=11), food preparation (n=10), medical (n=10), activity with others (n=8), and pet-related reminders (n=6). Overall, 38 reminders (45.2\%) resulted in correct code generation.

House chore reminders were the most common, followed by personal obligations. Within the house chore category, many participants (n=13) created reminders related to taking out the trash or completing routine household tasks. 
Across categories, nearly all reminders were mapped to time-based triggers (n=82 out of 84). Within these, one-time reminders were the most common (n=51). This indicates that even when users created their own reminders, they often relied on familiar time-based formulations rather than more complex trigger types.

\subsubsection{Prompted Reminder Creation Result}
The remaining scenarios (2–6) presented participants with a description of a situation and asked them to create a reminder based on it. These scenarios were designed to reflect situations in which users might naturally set reminders. By keeping the scenarios consistent across participants, we were able to examine how different users expressed reminder requests for the same situation.

\paragraph{\textbf{Scenario 2}}

Scenario 2 describes a situation where a reminder is needed to take food out of the microwave. Of the 39 times the scenario was presented, eight responses were user-generated scenarios, leaving 31 valid interactions. Across these interactions, time-based trigger were the most common (n=15), followed by state machine-based trigger (n=12), with other trigger types less common. In terms of reminder purpose, alerts were the most common reminder category (n=17), followed by one-time reminders and habits reminders.

Qualitatively, participants expressed this reminder in several different ways. Some used time-based triggers, either specifying a fixed time or a delay (e.g., ``Remind me to check the microwave 7 minutes from now''). Others (n=3) referred to the microwave itself, requiring sensor-based triggers. A larger group (n=12) combined these approaches, expressing multi-step logic that required a state machine trigger (e.g., ``Remind me when my food is done in the microwave, and then 3 minutes after it is done''). In addition, some participants (n=6) referenced external events such as a TV show mentioned in the prompt (e.g., ``Hello, I need a reminder to get my food out of the microwave during my favorite TV show''). Overall, this scenario showed the highest variety in both trigger types and expression styles.

System performance was the lowest in this scenario, with only 4 out of 31 interactions (12.9\%) resulting in correct code generation. 
A key pattern was that the system often prioritized the mention of the microwave and defaulted to sensor-based triggers, regardless of the participant’s intended reminder type. This bias also affected more complex requests. Although 12 interactions required state-machine logic (i.e., a combination of event-based and time-based conditions), the system generated it correctly in two cases and failed to recognize this need in the remaining instances (10 out of 12), instead defaulting to simpler sensor-based triggers.
The system also had a difficult interpreting time-based triggers, even when participants provided explicit timing in the first sentence of the interaction. Requests involving relative phrasing (e.g., ``before favorite show'') further increased ambiguity, due to the use of ``before'' and the absence of associated activity. In some cases, the system introduced unrelated activities, including ``relaxing'', ``meal preparation'', and ``eating'', that did not align with the participant’s intent. Overall, Scenario 2 showed lower consistency and accuracy compared to other scenarios.

\paragraph{\textbf{Scenario 3}}

Scenario 3 describes a situation where a user feels uneasy about whether the stove has been left on. Of the 35 completed instances, 10 were user-generated reminders, leaving 25 valid interactions. Across these interactions, time-based triggers were the most common (n=14), and habit was the most common reminder category (n=14), with other triggers and reminder categories less frequent.

Qualitatively, this scenario showed variation in both trigger types and expression strategies. Three of the four trigger types (time-based, activity-based, and sensor-based) and three categories (habit, one-time, and alert) were observed. 
Time-based triggers were the most common; however, in many cases participants shifted to time-based formulations after their initial requests were rejected by the system, rather than using them as a first strategy. Sensor-based triggers, particularly those tied to the stove, were the next most common approach. Unlike Scenario 2, where participants often expressed multi-step logic, requests in this scenario were typically simpler, such as reminders triggered when the stove was used. A consistent pattern across interactions was that participants initially requested reminders to turn off the stove ``before leaving the house,'' and then reformulated their requests when this phrasing was not supported. 

System performance in this scenario was moderate, with 12 out of 25 interactions (48.0\%) resulting in correct code generation. The scenario prompt led many participants to request reminders “before leaving the house.” However, the system rejected this trigger because it requires predicting the occurrence of an activity (``leaving\_home''), which was beyond its supported capabilities.
The system continued to reject the request even when participants provided additional timing information. In some cases, the system recovered after participants rephrased their requests, while in others the interaction ended prematurely. During code generation, the system frequently defaulted to sensor-based triggers due to the mention of the stove, even when the participant’s final request specified a different trigger type. Overall, the system showed inconsistencies in handling relative phrasing and often defaulted to sensor-based triggers, leading to errors in capturing user intent.

\paragraph{\textbf{Scenario 4}}
Scenario 4 describes a situation where a daughter wants to ensure that her mother remembers to eat the breakfast she prepared. Of the 40 times the scenario was presented, eight were user-generated reminders and 32 interactions were valid. Across these interactions, trigger types were primarily time-based (n=26), followed by activity-based (n=6). In terms of reminder category, most reminders were one-time reminders(n=29), with a smaller number of habit-based reminders (n=3). Notably, 23 out of 40 reminders were both time-based trigger and one-time reminders.

Despite limited variation in reminder types, participants showed variation in how they interpreted and expressed the task. A key difference was who the reminder was intended for: approximately half of the participants directed the reminder to the mother, while others framed it from the daughter’s perspective. Some participants also incorporated contextual cues from the prompt, such as sending the reminder when the mother wakes up to avoid disturbing her. However, most participants defaulted to time-based triggers, either by choice or by responding to the system’s follow-up questions.

System performance in this scenario was relatively low, with 7 out of 32 interactions (21.9\%) resulting in correct code generation. One issue was the incorrect mapping of contextual cues to activities. For example, references to ``breakfast'' were sometimes mapped to the ``Eating'' activity, which does not align with the intended reminder logic. In other cases, unrelated triggers (e.g., fridge or front door sensors) were introduced. The system also showed difficulty generating correct time-based logic, with 14 out of 23 time-based reminders resulting in incorrect code. The few successful cases (n=7) were typically those where participants used activity-based triggers, such as reminders tied to waking up, allowing the system to leverage the ``Sleeping'' activity. Overall, this scenario shows challenges in interpreting context involving multiple people and selecting appropriate triggers, leading to inconsistencies in how user intent was translated into reminder logic.

\paragraph{\textbf{Scenario 5}}

Scenario 5 describes a situation where a healthcare worker needs help remembering to wash her hands. Of the 40 completed instances, 10 were user-generated reminders, leaving 30 valid interactions. Across these interactions, activity-based triggers were the most common (n=19), followed by time-based triggers (n=11), and habit was the most common reminder category (n=29), with only one one-time reminder.

Qualitatively, the limited distribution of types and categories suggests a relatively consistent interpretation of the scenario. However, participants still varied in how they expressed reminder requests. The main difference was whether the reminder was triggered upon returning home after work (activity-based) or at a specific time. Within time-based triggers, most participants selected a single time, while some requested repeated reminders (e.g., hourly in the evening), indicating a need for repeated prompts to complete basic routine tasks. Within the activity-based triggers, participants differed in specificity, such as ``every time I arrive home from work'' versus ``every time I arrive home'' which changes the intent context and the constraints of the reminder.

System performance was highest in this scenario, with 23 out of 30 interactions (76.7\%) resulting in correct code generation. The system generally handled activity-based triggers well, successfully mapping them to available trigger context. However, it had difficulty incorporating additional constraints, such as specific times or conditions, such as returning home from work. Time-based triggers were also handled successfully in most cases. Overall, this scenario shows that simpler and more consistent reminder structures led to higher system performance, although incorporating additional constraints remained challenging.

\paragraph{\textbf{Scenario 6}}

Scenario 6 describes a situation where a senior, Margaret, needs to take supplements after dinner. Of the 39 completed instances, eight were user-generated reminders, leaving 31 valid interactions. Across these interactions, time-based triggers were slightly more common (n=17), followed by activity-based triggers (n=14), and habit was the most common reminder category (n=29), with only two one-time reminders.  

Qualitatively, this scenario showed low variation in expression. For the time-based triggers, participants either specified a time directly or provided one in response to the system’s follow-up questions. Most participants (n=13) selected times between 6:00 and 8:15 pm, while a few (n=3) chose other times despite the ``after dinner'' context. For activity-based triggers, participants typically expressed the reminder as occurring after dinner, which was mapped to available activity context.

System performance in this scenario was relatively low, with 22 out of 31 (71.0\%) interactions resulting in correct
code generation. Almost half of the time-based triggers were resulted in incorrect code, even though the interactions were simple and included explicit timing. In these cases, the system recognized the reminder as time-based but did not correctly incorporate the specified time. In contrast, activity-based triggers were generated more consistently, typically mapped to the ``Eating'' activity. However, the system did not distinguish between meals, treating all eating events equally, which does not fully capture the intended ``after dinner'' condition. Overall, this scenario shows that while simpler expressions reduced variability, the system still struggled to incorporate specific temporal constraints and finer-grained activity distinctions.

\subsection{From Study 1 to Study 2}
Study 1 demonstrated users were able to express their reminder intent in flexible and varied ways, demonstrating the potential of natural language and conversational interaction for reminder authoring. 
At the same time, it showed both the strengths and limitations of the initial reminder authoring pipeline.
We observed mismatches between how users described reminders and what the system could support, which sometimes led to incomplete or non-executable reminder logic. These mismatches were especially common when users used relative time expressions or described reminders with multiple steps or conditions. We also observed that many users relied on time-based triggers as a familiar starting point, suggesting that reliable support for time-based expressions is essential for usability. While the system handled some simple time-based triggers, more complex or relative time related expressions (e.g., ``later tonight'' or ``after dinner'') were not always interpreted consistently. 

These observations shows both the challenges and potential of natural language reminder authoring. The initial pipeline demonstrated that conversational interaction can effectively capture user intent, while also revealing opportunities to improve reasoning, feasibility guidance, and follow-up clarification. Based on these insights, we refined the authoring pipeline to better support how users naturally describe reminders and to improve the reliability of generating executable, context-aware reminders. Study 2 was designed to evaluate whether these refinements enable the system to better handle diverse user inputs while maintaining a smooth and understandable authoring experience.

\section{Study 2: Evaluating the Refined Authoring Pipeline}
\subsection{Updated System Overview}

\begin{figure*}[t!]
  \includegraphics[width=\textwidth]{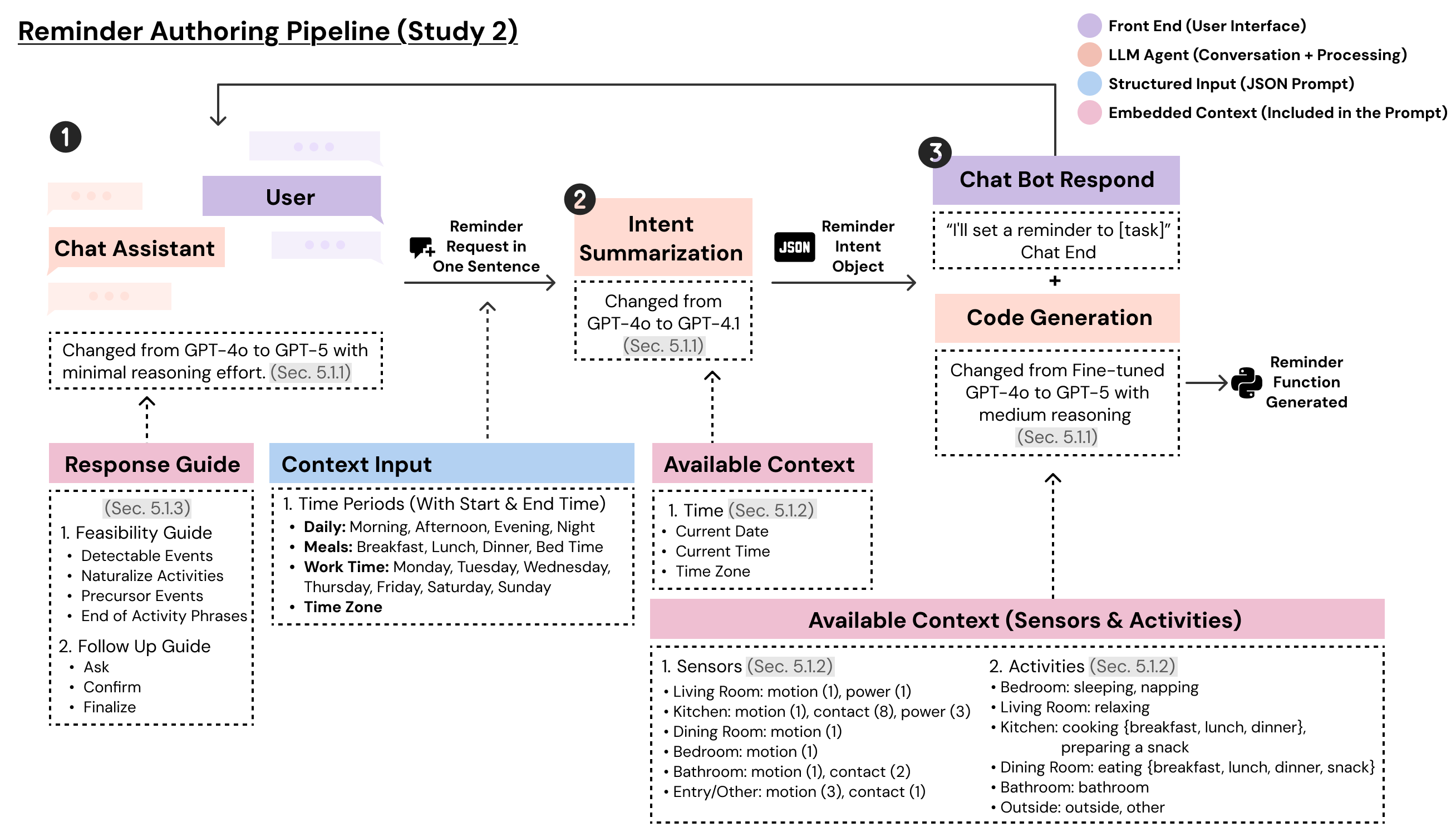}
  \caption{Overview of the updated reminder authoring pipeline used in Study 2. Components that were revised from Study 1 are highlighted in gray and annotated with corresponding section numbers.}
  \label{fig:system}
  \Description{Overview of the updated reminder authoring pipeline used in Study 2. Components that were revised from Study 1 are highlighted in gray and annotated with corresponding section numbers.}
\end{figure*}

In this section, we present the updated reminder authoring pipeline used in Study 2, which was revised based on findings from Study 1. Figure~\ref{fig:system} shows an overview of the updated system pipeline. Similar to the initial version, users create reminders in natural language through the same chat-based interface. Although the overall user workflow remains consistent with Study 1, the updated pipeline includes several revisions to the model choices, the way context is incorporated, and the structure of the processing pipeline. These updates are described in the following subsections.

\subsubsection{Model Changes}
Overall improvements in LLMs, such as the released of GPT-5, improved the pipeline's ability to follow instructions, understanding of time-related expressions, and reminder code generation, while maintaining appropriate response speed for the conversational interaction.

The \textbf{chat assistant} was changed from GPT-4o to GPT-5 with minimal reasoning effort. Based on findings from Study 1, GPT-4o often failed to follow instructions, struggled to infer details from earlier turns in the conversation, and unnecessarily repeated questions. For example, the assistant sometimes asked whether a reminder should occur ``just this once, or every time'' even when the user had already indicated that it was a routine reminder. GPT-5 was able to handle more complex instructions and infer missing details from the broader conversational context.

The \textbf{intent summarization} model was changed from GPT-4o to GPT-4.1. In Study 1, GPT-4o showed inconsistencies in resolving relative time related expressions such as ``5 minutes later'' and occasionally failed to assign appropriate times to parts of the day, such as night. Although we also considered GPT-5 for this stage, we selected GPT-4.1 because it performed better than GPT-4o in relative time interpretation and was faster and less costly than GPT-5. This was important because the reminder authoring pipeline involves multiple sequential LLM calls, and increased latency at any stage affects the system's overall responsiveness.

The \textbf{code generation} model was also updated, moving from a fine-tuned GPT-4o model trained on 23 reminders to GPT-5 with medium reasoning. In Study 1, the fine-tuned GPT-4o model often produced code that was syntactically correct but logically incomplete, particularly for reminders involving delays, multi-step conditions, or state-dependent logic. For instance, in one case, the generated code correctly detected the end of sleep but failed to implement the user-requested one-hour delay before the reminder should fire. These failures showed that code generation in this setting requires reasoning over time, state, and conditional structure. The updated model will improve the consistent handling of delays, multi-step logic, activities, and compound conditions.

\subsubsection{Context Incorporation Changes}
Based on findings from Study 1, we revised how contextual information is incorporated into the pipeline.
In Study 2, critical contextual variables were embedded more directly into the LLM prompts instead of simply appending the context. This change improved consistency, reduced pipeline complexity, and supported more reliable reasoning.

For intent summarization, Study 1 revealed inconsistencies in resolving relative time expressions and mapping vague temporal phrases to concrete values. To address this, we embedded the current date (e.g., \texttt{CURRENT\_DATE: \{current\_date\}}) and predefined time mappings (e.g., \texttt{TIME\_MAPPINGS: \{time\_mappings\}}) directly into the prompt. These inputs provided structured context for interpreting expressions such as ``today'' or ``after dinner,'' improving the consistency of temporal resolution without relying on external processing steps.

The assistant was also provided with an updated set of detectable events embedded directly in the prompt. In Study 1, sensors and activities were represented as flat labels. In Study 2, we extended this representation to include additional context, such as the locations and types of sensors, as well as the locations associated with activities. This change provides more structured contextual information and enables the assistant to better ground event-based reminder conditions.

For code generation, Study 1 showed that the model often omitted time delays or incompletely represented multi-step conditions. To address this, we updated the prompt to incorporate runtime context, including the current time, and to require explicit handling of delays when specified in the user’s request. Additional rules were introduced to improve consistency in interpreting activity-based conditions, such as treating ``after [activity]'' as referring to the end of the activity. These changes improved the system’s ability to generate code that captures delays.

\subsubsection{Processing Pipeline Changes}

Based on the findings from Study 1, we revised the structure of the reminder-authoring pipeline used in Study 2. In Study 1, the feasibility check was evaluated as a separate step after the conversation ended. As a result, users often completed the full authoring interaction before being informed that their requested reminder was not supported. In addition, feasibility checks were not always consistent. These limitations disrupted the flow of interaction and reduced the effectiveness of the authoring process.

In Study 2, we replaced this feasibility check with an in-conversation approach. Instead of evaluating feasibility after the reminder was fully specified, the system was designed to guide users toward feasible configurations during the interaction. To support this, we embedded a structured feasibility guide directly into the chat assistant’s prompt. This guide included a predefined set of detectable events (e.g., ``when you arrive home’’), common activity labels (e.g., sleeping, napping), and mappings between activities and relevant events (e.g., ``morning tasks’’ mapped to ``when you wake up’’), as well as phrases representing the completion of activities (e.g., ``when you finish eating’’). These elements provided the assistant with structured knowledge of detectable events and activities, enabling it to guide users toward reminder configurations that the system could support.

In addition, we introduced structured guidelines for follow-up questions to improve the system’s ability to identify and request missing information. In Study 1, the assistant often asked redundant or unnecessary follow-up questions, even when the required information had already been provided. To address this, we introduced a structured interaction pattern consisting of three stages: ask, confirm, and finalize. In the ask stage, the assistant was instructed to request only the single most important missing piece of information, such as whether a reminder should be triggered by an event or a specific time. In the confirm stage, once the trigger type had been identified, the assistant restated the reminder in either event-based or time-based form and asked the user to verify that interpretation. In the finalize stage, after receiving confirmation, the assistant produced a final summary of the reminder and ended the interaction. This interaction structure was intended to reduce redundant questioning, improve clarity, and make the authoring process more predictable for users.

\subsection{Study Design}
Study 2 was conducted to evaluate improvements to the reminder authoring pipeline identified from Study 1. The study followed the same procedure as Study 1, with the interface and reminder-creation scenarios left unchanged. Given the positive usability results observed in Study 1, we retained the same interface to isolate the impact of the refined pipeline while maintaining the same user interaction conditions. Participants for this follow-up study were also recruited through Prolific and received \$6 USD for completing the study. Prolific participants who completed Study 1 was excluded. Other recruitment criteria and study approval followed the same procedure as Study 1.

\subsection{Result}

\subsubsection{Participant Demographic Information}
In this study, 10 participants (5 male, 4 female,  and 1 non-binary) were recruited. Participants ranged in age from 26 to 53 years ($M = 40.3$, $SD = 9.42$). Participants reported varied LLM usage frequency ($M = 4.5$, $SD = 2.12$ on a 7-point scale, where 7 indicates daily usage), with responses spanning from no prior use to frequent daily use.

\subsubsection{System Usability Scale (SUS) Results}
In this study, the revised interface received a mean SUS score of 85.00 (SD = 12.19, $n=10$). This increase from $79.81$ is consistent with our iterative improvements to the chat assistant guidance and the system's ability to surface relevant sensors and events.

\begin{figure*}
  \includegraphics[width=\textwidth]{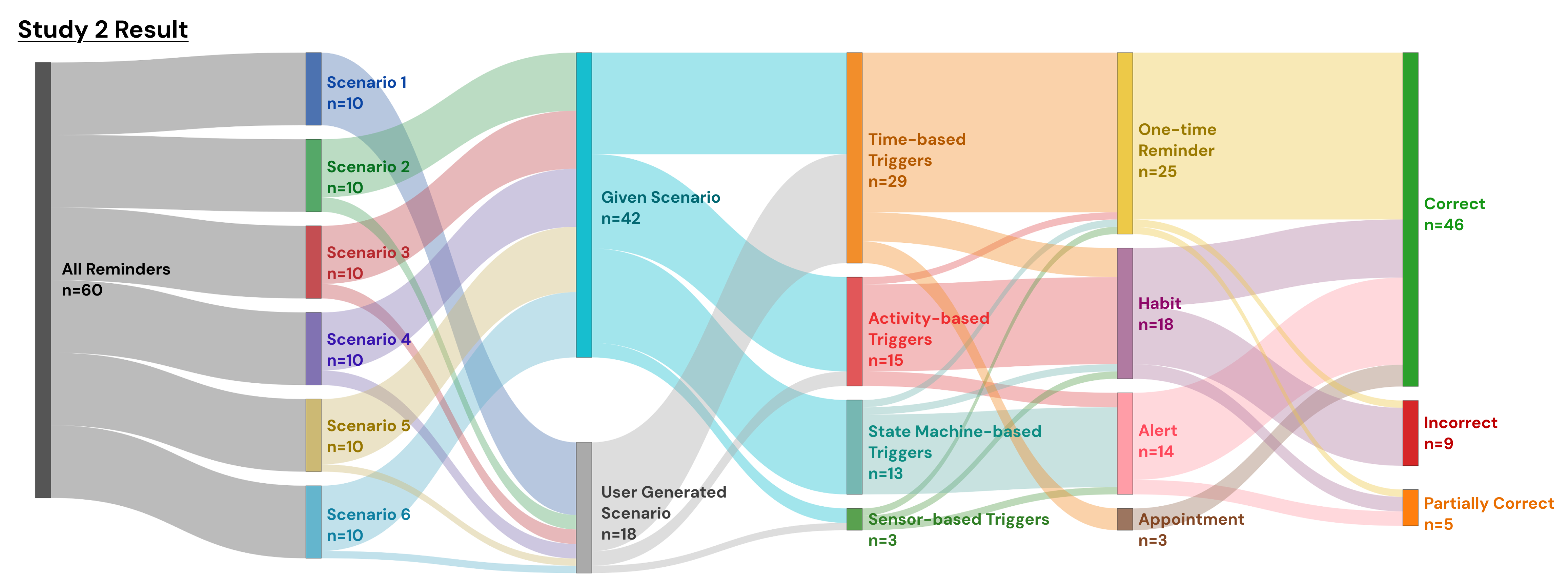}
  \caption{Study 2 results overview. Sankey diagram illustrating the distribution of results across categories. The diagram follows the same structure as Study 1 but omits the validation stage, as all reminders are valid. Reminders are first grouped by scenario and then categorized as either given or user-generated, before being mapped to reminder trigger types and reminder purpose categories. Final outcomes reflect the correctness of the generated code.}
  \label{fig:study2_result}
  \Description{Study 2 results overview. Sankey diagram illustrating the distribution of results across categories. The diagram follows the same structure as Study 1 but omits the validation stage, as all reminders are valid. Reminders are first grouped by scenario and then categorized as either given or user-generated, before being mapped to reminder trigger types and reminder purpose categories. Final outcomes reflect the correctness of the generated code.}
\end{figure*}

\subsubsection{Overall Scenarios Result}

In Study 2, 10 participants each completed 6 reminder-authoring scenarios, resulting in 60 reminder-creation sessions. No sessions reached the attempt limit, and none were excluded from analysis. We analyzed all 60 sessions using the same qualitative content analysis method with an iterative categorization scheme. Participants created reminders based on the same six predefined scenarios (Scenarios 1–6). Similar to Study 1, some participants did not follow the given instructions and instead created their own reminders. As in Study 1, these were labeled as \textit{user-generated scenarios}. Figure~\ref{fig:study2_result} illustrates the distribution of reminders across scenario type, reminder type, category, and code generation outcomes. The distribution of the result is also shown in Table~\ref{tab:study1_study2_scenarios}. On average, participant took $3.28$ turns ($SD = 1.19$, $n = 60$) when constructing the reminders (shown in Figure~\ref{fig:number_of_turns}). 

Across all sessions, time-based triggers were the most common (n=29), followed by activity-based triggers (n=15) and state machine-based triggers (n=13). Sensor-based triggers were less common (n=3). Reminders were further categorized into four reminder groups, consistent with Study 1: one-time reminders (n=25), habits (n=18), alerts (n=14), and appointments (n=3). In the following, we provide a more detailed analysis of \textbf{user-generated scenarios} (n=18) and \textbf{prompted scenarios} (n=42), focusing on category-specific distributions.

\subsubsection{User-Generated Scenario Reminder Creation}

In Study 2, a total of 18 user-generated reminders were collected across scenarios, including 10 from Scenario 1 and seven from other scenarios where participants chose to create their own reminders instead of following the given instructions. The distribution of the reminder type for this user-generated scenario is shown in the Table~\ref{tab:reminder_types}. These reminders covered several categories, including medical (n=5), personal obligations (n=4), pet-related reminders (n=4), house chores (n=2), food preparation (n=2), and activity with others (n=1).

The correct code generation rate improved substantially compared to Study 1, increasing from 45.2\% to 77.8\% (14 out of 18 reminders). This improvement is consistent with performance gains observed across structured scenarios, suggesting that improvements to the system’s code generation generalized to user-generated inputs. Similar to Study 1, most user-generated reminders used time-based triggers and one-time reminders, which were generally handled successfully. The few incorrect code outputs were distributed across different categories without a clear pattern, making it difficult to attribute failures to specific reminder types. Overall, the high success rate indicates that the system can reliably handle varied reminder requests across different domains.

\subsubsection{Prompted Reminder Creation}
The same scenarios (2–6) were used in Study 2, following the design described in Study 1. This allowed us to examine how changes in the system influenced reminder authoring while keeping the task conditions consistent across studies.

\paragraph{\textbf{Scenario 2}}
In Scenario 2, of the 10 completed instances, nine were valid interactions and one was a user-generated scenario. Across these interactions, state-machine triggers were the most common trigger (n=7), and alert was the dominant reminder category (n=7), with other triggers and categories less frequent. System performance in this scenario improved substantially compared to Study 1, with 7 out of 9 interactions (77.8\%) resulting in correct code generation, compared to 4 out of 31 (12.9\%) in Study 1. This improvement was most evident for state-machine triggers, with 6 out of 7 correctly generated in Study 2, compared to 2 out of 12 in Study 1.

This improvement can be attributed to changes in the system’s interaction design. In Study 1, the system often defaulted to sensor-based triggers when the microwave was mentioned and frequently failed to recognize when state-machine logic was required. In Study 2, the system used more targeted follow-up questions (e.g., ``Should I remind you when the microwave finishes, or at a specific time?''), which guided participants toward supported trigger types and made system capabilities more explicit. Overall, targeted follow-up questions improved the system’s ability to identify appropriate trigger structures, leading to more accurate code generation for complex reminder types.

\paragraph{\textbf{Scenario 3}}

In Scenario 3, of the 10 completed instances, eight were valid interactions and two were user-generated scenarios. Across these interactions, state-machine triggers (n=4) and activity-based triggers (n=2) were the most common types, and alert was the dominant reminder category (n=8), reflecting participants framing the reminder as a safety check tied to leaving the house. 
System performance improved substantially compared to Study 1, with approximately 7 out of 8 interactions (87.5\%) resulting in correct code generation, compared to 12 out of 25 (48.0\%) in Study 1.

This improvement can be attributed to changes in the system’s handling of leaving-home triggers. In Study 1, many interactions (n=19) began with requests for reminders before or upon leaving the house, which the system often rejected, forcing participants to reformulate their requests as time-based triggers. This contributed to the dominance of time-based triggers and the system's tendency to default to stove sensor as triggers. In Study 2, this issue was largely resolved, allowing participants' original requests to succeed rather than requiring repeated reformulation. Overall, the updated pipeline improved trigger handling, reduced the need for request reformulation, and supported more accurate code generation.

\paragraph{\textbf{Scenario 4}}

In Scenario 4, of the 10 completed instances, eight were valid interactions and two were user-generated scenarios. Across these interactions, time-based trigger and one-time reminders were the most common (n=7), with one state-machine reminder. This distribution is consistent with Study 1, where time-based one-time reminders were also dominant. System performance improved substantially, from 7 out of 32 interactions (21.9\%) in Study 1 to 8 out of 8 (100.0\%) in Study 2.

This improvement can be attributed to the updated context incorporation and structured follow-up approach. In Study 1, the system often defaulted to incorrect activity mappings (e.g., triggering on ``eating'' when breakfast was mentioned), or producing unrelated triggers such as the fridge or front door opening. It also had difficulty handling the multi-person framing of the scenario. In Study 2, these issues were largely reduced, with the system asking more targeted follow-up questions and mapping the reminder to appropriate triggers without introducing irrelevant options. Overall, the updated pipeline improved handling of context and reduced common failure patterns, leading to more consistent code generation in this scenario.

\paragraph{\textbf{Scenario 5}}

In Scenario 5, of the 10 completed instances, eight were valid interactions and two were user-generated scenarios. Across these interactions, activity-based triggers were the most common type (n=5), and habit was the dominant reminder category (n=7), consistent with Study 1, with a state-machine (n=1) and sensor-based (n=1) triggers also appeared. System performance was approximately 6 out of 8 (75.0\%), comparable to Study 1's 23 out of 30 (76.7\%), making this the scenario with the least change between the two studies.

This consistency reflects that activity-based habit reminders triggered by arriving home were already well-supported in Study 1.  The two incorrect interactions highlight a remaining challenge in handling more specific conditions. In particular, when participants specified ``after returning home from work'' rather than simply ``when I arrive home,'' the system did not always incorporate the additional constraint. Overall, the system maintained strong performance in this scenario, with limited change following the pipeline updates.

\paragraph{\textbf{Scenario 6}}

In Scenario 6, of the 10 completed instances, nine were valid interactions and one was a user-generated scenario. Across these interactions, activity-based (n=5) and time-based (n=4) triggers were similarly common, and habit was the dominant reminder category (n=8), a distribution similar to Study 1. System performance was 4 out of 9 (44.4\%), representing a decline from Study 1, with 21 out of 31 (67.7\%) interactions.

This result reflects challenges in handling activity specificity. For activity-based triggers, the system did not always distinguish between “dinner” and general eating events, and often mapped requests to broader activity categories. In addition, when participants used phrases such as ``after dinner,'' the system sometimes interpreted these as location-based triggers (e.g., entering the dining room), rather than capturing the intended time-related relationship. These patterns suggest that while the updated pipeline supports structured activity and location information, finer-grained distinctions in meal-related language remain more difficult to represent. Overall, this scenario highlights an area where system improvements are still needed, particularly in capturing nuanced activity semantics in everyday language.

\section{Discussion}

\begin{table}[t!]
\centering
\footnotesize
\renewcommand{\arraystretch}{1.4}
\begin{tabular}{p{1.5cm} p{13cm}}
\toprule
\textbf{Scenario} & \textbf{Variation of Instructions} \\
\midrule
Scenario 2 & Please create a reminder to check the stove when I leave the house. (S1-P1)\\
           & \cellcolor{rowgray} 5 minutes after I start the microwave, please remind me to go check on my food. (S1-P20) \\
           & Hello, I need a reminder to get my food out of the microwave during my favorite TV show. (S1-P22) \\
           & \cellcolor{rowgray} Please remind me to take my food out in 5 minutes. (S2-P1)\\      
\midrule
Scenario 3 & Can you set a reminder for me to remember to check if I properly turned my stove off? Everytime I use the stove I want this reminder. (S1-P9)\\
           & \cellcolor{rowgray} Can you remind me to turn off the stove in the morning. (S1-P12)\\
           & I love cooking in the morning but sometimes I tend to forget to turn the stove off. Can you please provide me a reminder? (S1-P17) \\
           & \cellcolor{rowgray} I would like to set a reminder to remember to turn the stove off after cooking breakfast. (S2-P2)\\
\midrule
Scenario 4 & Please tell my mother there is breakfast in the fridge for her when she starts the coffee pot in the morning. (S1-P20)\\
           & \cellcolor{rowgray} Remind my mother breakfast is in the refrigerator at 10:00 AM. (S2-P3)\\
           & Remind me to text mom tomorrow morning at 8:00 am to tell her to eat the breakfast I prepared for her. (S2-P4)\\
           & \cellcolor{rowgray} Can you please remind mother that I left food in the fridge for her when she wakes up? could you also reminder her again in case she forgets and never checks on it? (S2-P7) \\
\midrule
Scenario 5 & \cellcolor{rowgray} Please set a reminder to wash my hands after work, I get off around 7. (S1-P1)\\
           & I would like to set a reminder to wash my hands when I get home after work. (S2-P2)\\
           & \cellcolor{rowgray} Remind me to wash hands after work each day, 10 minutes after I open the door in the afternoon for the first time each day. (S2-P3) \\
           & Remind me when I arrive home to wash my hands when I open the front door. (S2-P6)\\
\midrule
Scenario 6 & \cellcolor{rowgray}  Remind me to take my supplements in the evening after dinner. (S1-P3)\\
           & Remind me to take medication at 6 PM every night. (S2-P3)\\
           & \cellcolor{rowgray} Always remind me around bedtime to take my supplements. (S2-P10) \\
           & I forget to take supplements sometimes. Remind me to take them. (S2-P14)\\
\bottomrule
\end{tabular}
\caption{Example variations in how participants expressed reminder intent across scenarios. Labels in parentheses indicate the study (S1 or S2) and participant number (P).}
\label{tab:reminder_variation}
\end{table}

\subsection{Diversity in Reminder Authoring Expressions}
Our findings show that users often use diverse everyday language forms to describe the same reminder scenario, varying in specificity, structure, and trigger type (examples are shown in Table~\ref{tab:reminder_variation}). For example, some participants expressed reminders using explicit times, while others relied on event-based or routine-based expressions (e.g., ``after dinner'' or ``when I get home'') for the same task. In addition, users differed in how they structured reminder logic, with some specifying simple triggers and others expressing multi-step or conditional sequences (e.g., ``remind me if I haven’t done [X] after [Y]''). 

We also observed consistent differences in how participants interpreted and prioritized information within the same prompt. Some participants consistently provided minimal input, focusing only on the core task.  For example, one participant simply typed ``To put my stove off'', not even a full sentence was provided,  while others included additional context or background (e.g., ``I always get distracted after using the microwave and forget about it''). Others adopted a more structured approach, providing detailed and complete instructions upfront, or consistently initiating interactions with conversational elements such as greetings. These patterns suggest that users not only vary in \textit{what} they express, but also in \textit{how} they approach the task of reminder authoring.

This aligns with prior work in end-user programming, which shows that users vary widely in how they conceptualize and express tasks. It also highlights a key limitation of traditional rule-based systems such as using TAP, which require users to translate their intent into predefined structures, constraining how reminders can be expressed.

Taken together, these observations from our study support our vision where some reminder authoring cannot be effectively supported by a single predefined representation or interaction pattern. Systems that enforce fixed structures may fail to capture the range of user intent, particularly when expressions differ in abstraction, completeness, or structure. 
Our function representation enables the expression of more complex and varied reminder logic.
These findings point to a broader design implication: reminder authoring systems need to support the multiple valid interpretations of the same task and adapt to different user styles of expression. More importantly, systems should treat reminder authoring as a collaborative process, where user input is interpreted, clarified, and refined over time, rather than as a one-step specification task.

\subsection{Handling Vague and Under-specified Reminder Requests}
Reminder requests were sometimes under-specified, with participants providing less information than required to construct a complete reminder. For example, some users issued minimal requests such as ``turn off the stove,'' without specifying when or under what conditions the reminder should occur. 
Others used expressions such as ``around my bedtime'' which required the system to interpret when certain events typically occur.
The challenge therefore lies not only in handling such expressions, but also in cases where key components of the reminder are missing entirely. Prior systems, such as the LLM-powered smart home assistant Sasha~\cite{king2024sasha}, have explored generating automation routines from everyday language. These approaches attempt to handle under-specified requests by inferring missing triggers and actions to create a complete executable plan upfront, followed by user critique. However, this can lead to incorrect or undesired outcomes when the system's inferences do not align with user intent. Our findings highlight the importance of conversational interaction in reminder authoring. Rather than fully specifying their requests upfront, participants often relied on the system to guide them toward a complete and actionable reminder. In our system, conversational interaction functions as a collaborative process for clarifying missing intent. 

Our observations suggest a key design implication: handling under-specified input cannot rely on a one-step interpretation but instead requires incremental clarification through follow-up questions, allowing users and systems to collaboratively construct complete reminders. Instead of inferring and guessing missing details in a single step, reminder authoring systems should guide users toward specifying only the necessary information needed to produce executable and context-appropriate outcomes.

\begin{figure*}
  \includegraphics[width=\textwidth]{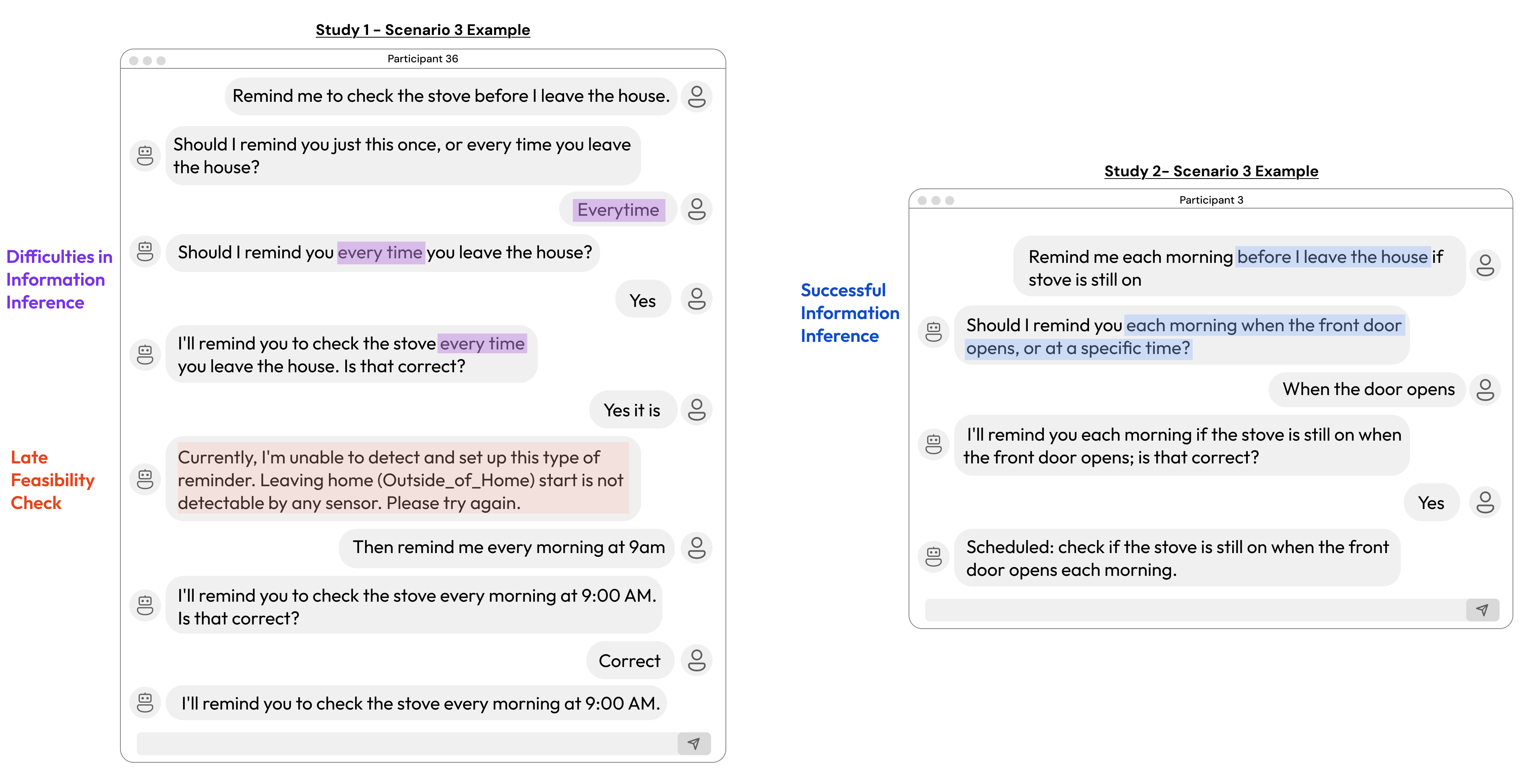}
  \caption{Comparison of reminder authoring interactions in Scenario 3 between Study 1 (left) and Study 2 (right). In Study 1, the system struggles to infer the intended trigger and introduces a late feasibility check, leading the user to reformulate the request as a time-based reminder. In Study 2, the system uses targeted follow-up questions to clarify the trigger early in the interaction, enabling successful interpretation of the intended reminder.}
  \label{fig:chat_example}
  \Description{Comparison of reminder authoring interactions in Scenario 3 between Study 1 (left) and Study 2 (right). In Study 1, the system struggles to infer the intended trigger and introduces a late feasibility check, leading the user to reformulate the request as a time-based reminder. In Study 2, the system uses targeted follow-up questions to clarify the trigger early in the interaction, enabling successful interpretation of the intended reminder.}
\end{figure*}

\subsection{The Role of Follow-up Questions in Reminder Authoring}
Follow-up questions play a central role in how users complete reminder requests. When the user’s initial reminder request is underspecified, especially in specifying when the reminder should occur, the system’s response directly shapes how users provide final reminder request. The wording of that question can influence users toward different types of triggers. 

Well-structured follow-up questions can improve clarity and reduce user effort. In our system pipeline in Study 1, the system consistently follows up by asking, ``Should I remind [task] just this once, or every time?'' prompts users to specify reminder frequency in a simple and understandable way. This type of question helps resolve ambiguity quickly and lowers the effort required to complete the interaction. However, this simplicity also introduces limitations. By presenting recurrence as a binary choice, the system implicitly constrained how users conceptualized reminder frequency. As a result, users often framed their requests within these options, even when their needs were more nuanced (e.g., conditional or context-dependent recurrence). In Study 2, the updated system pipeline enabled more targeted follow-up questions that better reflected available trigger types. Instead of narrowing user input prematurely, the system offered alternatives that preserved multiple interpretations (e.g., asking whether a reminder should occur at a specific time or when an event happens). This shift supported more accurate mapping between user intent and system capabilities. Figure~\ref{fig:chat_example} illustrates this contrast, showing how targeted follow-up questions in Study 2 support earlier clarification and successful reminder creation compared to Study 1.

System feedback further influences how users adapt their requests. In Study 1, feasibility feedback was introduced late in the interaction, with the system rejecting a reminder only after the conversation appeared complete and the reminder had been confirmed.
When the system indicated that a reminder type was not supported (e.g., "Currently, I'm unable to detect and set up this type of reminder"), participants often abandoned their original intent. Without guidance on how to revise their request, participants often reformulated their request as a time-based trigger instead of attempting to clarify or re-express the original trigger (Figure~\ref{fig:chat_example}). This behavior suggests that users tend to accommodate system constraints rather than challenge them, even when doing so results in a less suitable reminder. 

These observations highlight a key design implication: follow-up interactions should guide users without constraining their intent. Systems should ask targeted questions that clarify missing information and provide sufficient guidance while preserving multiple possible interpretations. This enabling users to refine their requests without being pushed toward simplified or system-preferred solutions.

\subsection{Exploring Opportunities in Reminder Authoring}
Smart home sensing creates opportunities for reminder authoring that go beyond fixed time or singular reminders. Because our system pipeline has access to contextual information such as smart home devices, activity labels, time of day, and household routines, it can support reminders tied to events and situations rather than only clock times. For example, tasks such as checking the laundry, taking out the trash, or locking the door may be more meaningfully connected to household states or routines than to a fixed schedule.

However, our studies suggest that while the system enables richer forms of reminder authoring, users are still learning how to take advantage of these capabilities. Participants often defaulted to familiar time-based triggers, even when the scenario could support event-based or context-aware triggers. In other cases, users expressed intent in ways that were meaningful to them but did not directly map to the system’s available trigger types. For example, some users requested reminders ``before'' an event. These patterns indicate a gap between what the system can support and how users understand and access those capabilities, highlighting an opportunity for capability discovery during the reminder authoring process. Rather than waiting for users to know which triggers are supported, the system can help users discover feasible options during the conversation. For example, when a user asks for a reminder related to dinner, the system could suggest alternatives such as a reminder after eating or a reminder after a kitchen activity, depending on what the home can detect. Similarly, when a user mentions an appliance, the system could suggest whether the reminder should depend on time, device usage, or a sensor event.
For example, instead of detecting whether someone has left the home, the system can suggest using the opening of the front door as a proxy. With improved context integration, the chat assistant can proactively suggest alternatives in some cases.

The broader design implication is that the reminder authoring system should help users learn and use system capabilities without requiring them to understand the underlying sensors or activity models. Context-aware reminder systems should not only execute supported reminders, but also make the space of possible reminders visible during authoring. In this way, the system can help bridge user intent with what is possible, helping users create reminders that suit their daily lives.

\subsection{Transparency and User Control in Context-Aware Reminder Authoring}
The smart home context should be treated as a resource to assist with reminder authoring rather than as a substitute for user intent. In our system, the chatbot-style user interface is designed to support this balance by providing an interaction format that allows users to articulate their needs, review system suggestions, and revise reminder configurations as needed. This interface preserves user agency while still allowing the system to leverage contextual information from the home.
The use of a conversational interface also makes contextual reasoning more interpretable and transparent. Because the system has all the resources to present its reasoning through dialogue, users can better understand how available home context informs reminder setup and why a reminder is configured in a particular way. Rather than functioning as a black box, the system can explain how contextual cues contribute to a context-aware reminder and allow users to confirm, modify, or reject those inferences. In this way, the interface provides support in a balance between contextual intelligence, flexibility, and user control.

\section{Limitation and Future Work}
This paper focuses on the authoring of personalized, context-aware smart home reminders: how users express reminder intent in everyday language and how a conversational system can translate that intent into structured, executable logic. 
Our contribution centers on the authoring process and our evaluation was conducted through online scenario-based studies rather than through deployment in real homes. This allowed us to compare how different users authored reminders across the same scenarios and to identify breakdowns in the pipeline. However, it does not capture the full complexity of real smart home use, including noisy sensor data, changing routines, and long-term adaptation. Future work should deploy the system in real homes to understand how context-aware reminder authoring works with real sensor streams and everyday household routines.

Our evaluation also focuses on reminder creation rather than the full lifecycle of reminder use. We studied whether users could create reminders and whether the system could translate those requests into executable logic, but we did not evaluate how reminders are delivered, reviewed, modified, or debugged after creation. In practice, users may need to understand why a reminder triggered, why it failed to trigger, or how to adjust it after experiencing it in daily life. Future work should extend the conversational interface to support post-authoring management, including explanation, editing, and repair of reminders in real scenarios.

Finally, the reminders our system can support are limited by the sensing and activity recognition capabilities available in the home. Our pipeline can help users map everyday language to feasible smart home triggers, but it cannot support activities or contextual distinctions that the underlying sensors and activity labels cannot detect. For example, the system still struggled with reminders such as “after dinner” when the available activity label only captured eating more generally. Future work should explore personalized activity models, user-defined activity labels, and open-vocabulary activity recognition to support a broader range of everyday reminder expressions.

\section{Conclusion}
In this paper, we presented a conversational pipeline for authoring personalized, context-aware smart home reminders using everyday language. The system translates natural language reminder requests into structured representations and executable logic that can incorporate time-based, activity-based, sensor-based, and state-machine-based triggers. Across two studies, we examined how users express reminder intent, where mismatches arise between user expectations and system capabilities, and how conversational guidance can help users create feasible reminders.

Our findings show that users describe reminders in diverse and sometimes under-specified ways. While many users rely on familiar time-based reminders, they also express needs that depend on routines, activities, device states, and sequences of events. The updated pipeline improved the system’s ability to handle these expressions by incorporating stronger conversational guidance, richer contextual information, and improved code generation. At the same time, our results show that context-aware reminder authoring is not only a technical translation problem. It is also an interaction problem that requires clarification, transparency, and user control.

Our work suggests that smart home contexts can serve as a valuable resource for reminder authoring, helping users create reminders that better align with their everyday routines. Rather than requiring users to manually configure sensors or write trigger-action rules, conversational systems can help bridge the gap between natural language intent and executable smart home logic. More broadly, our work points toward reminder systems that move beyond fixed times and simple triggers, enabling users to express reminders in the same flexible language they use to describe daily life.

\begin{acks}
This work was partially supported by NSF IIS-2112633. We thank our participants for their participation and our labmates for their helpful feedback.
Generative AI was used to improve the quality of writing, including style, phrasing, and grammar in this paper.
\end{acks}

\bibliographystyle{ACM-Reference-Format}
\bibliography{main_reference}

\end{document}